\documentclass[preprint,showpacs,showkeys,preprintnumbers,amsmath,amssymb]{revtex4-1}
\usepackage[utf8]{inputenc}

\usepackage{graphicx,subfigure}
\usepackage{psfrag}
\usepackage{amsmath}
\usepackage{bbm}
\usepackage{dsfont}
\usepackage{braket}
\usepackage{bbold}
\usepackage[pdfpagelabels=true]{hyperref}

\begin{document}


\title{Geometry of Gaussian quantum states}

\author{Valentin Link}
\author{Walter T. Strunz}
     \affiliation{Institut f\"{u}r Theoretische Physik,
Technische Universit\"at Dresden, D-01062 Dresden, Germany}


\email{walter.strunz@tu-dresden.de}

\date{\today}

\begin{abstract}
We study the Hilbert-Schmidt measure on the manifold of mixed Gaussian states in multi mode continuous variable quantum systems. 
An analytical expression for the Hilbert-Schmidt volume element is derived. Its corresponding probability measure can be 
used to study {\it typical} properties of Gaussian states. It turns out that although the manifold of 
Gaussian states is unbounded, an ensemble of Gaussian states distributed according to this measure still has a normalizable distribution of symplectic eigenvalues, from which
unitarily invariant properties can be obtained. By contrast, we find that for an ensemble of one-mode Gaussian states based on the Bures measure
the corresponding distribution cannot be normalized.
As important applications, we determine the distribution and the mean value of von Neumann entropy 
and purity for the Hilbert-Schmidt measure.
\end{abstract}

\pacs{03.67.-a, 42.50.Dv, 89.70.+c}
\keywords{Gaussian state, squeezed thermal state, Hilbert-Schmidt metric, Bures metric, entanglement, separability, geometry}

\maketitle

\section{Introduction}
The analysis of spaces of quantum states equipped with some physically motivated measure has been a lively field of research in the last decade.
These investigations help to shed light on "typical`` properties of these spaces and their elements, such as the purity, or the entropy of the states. 
In the case of multi-partite states, an understanding of metric properties allows for a
quantitative characterization of entanglement \cite{geometryQS}.
Detailed knowledge of the distribution of a class of states may also serve to specify a priori probabilities 
for quantum state estimation \cite{ZHradil} and for quantum tomography \cite{tomography}, and contributes to the understanding of the concept of ''unknown quantum
state`` \cite{bayes}. 

Most results in this field are restricted to quantum systems with a finite dimensional Hilbert space, such as many qubit systems, 
and cannot be easily generalized to systems with infinite dimensional Hilbert space. The important continuous variable systems belong to the latter and contain 
the class of mixed Gaussian quantum states, which will be considered in this article. Gaussian states are quantum states that have a Gaussian Wigner function, 
and they can be characterized by their covariance matrix alone \cite{AdessoAndvancesPerspectives}. These states are fundamental for
continuous variable quantum information science \cite{WeedbrookPirandola}.

For pure states there exists a unique unitarily invariant measure, the Fubini-Study measure that is induced by the Haar measure on the unitary group. 
This measure on the pure Gaussian states has already 
been studied in \cite{InvariantMeasuresPureGaussian}. In our work presented here, the Hilbert-Schmidt measure for {\it mixed} Gaussian states will be 
examined, because it is one of the most widely used invariant measures for mixed quantum states \cite{HallRandQC,Simon,PurityQubit}. 
To this aim a brief introduction to Gaussian quantum states is given in section \ref{sec:Gaussst}. The Hilbert-Schmidt volume element for Gaussian states 
is derived in section \ref{sec:HSvol}, and statistical properties of states distributed according to this measure are calculated in section \ref{sec:HSdist} 
and \ref{sec:HSprop}. In particular, the distribution of symplectic eigenvalues is calculated. These values can be used to determine all unitarily invariant 
properties of Gaussian states, such as the von Neumann entropy and the purity. In the following section \ref{sec:BURES} these results are compared to the 
Bures measure for the simple one mode case, which leads to qualitatively different results. Finally, in section \ref{sec:SUM} a summary of the results and an 
outlook for further research is given. 

\section{Gaussian quantum states}\label{sec:Gaussst}

Consider $N$ Fock spaces $\mathcal{H}_i$ in each of which one bosonic creation operator $a_i^\dagger$ is defined and an 
orthonormal basis is given by the number states $\{\ket{n}_i|n\in\mathbb{N}_0\}$. An $N$-mode continuous variable system is a 
quantum system described by the Hilbert space $\mathcal{H}=\bigotimes_{i=1}^{N}\mathcal{H}_i$, arising from the direct product 
of these Fock spaces. This space is spanned by the product basis 
$\{\ket{n_1,\hdots,n_N}=\ket{n_1}_1\otimes\hdots\otimes\ket{n_N}_N|n_1,\hdots,n_N\in\mathbb{N}_0\}$.\\
The characteristic function of a state $\rho$ in $\mathcal{H}$ is defined as:
\begin{equation}
 \chi({\alpha})=\text{tr}(\rho D_{{\alpha}}),\qquad D_{{\alpha}}=\prod\limits_{i=1}^N \exp{(\alpha_ia_i^\dagger-\alpha_i^*a_i)}\qquad \alpha\in\mathbb{C}^N\,.
\end{equation}
The complex Fourier-transform of the characteristic function is called the Wigner quasiprobability distribution in phase space \cite{introductoryQOgerry}.
A Gaussian state in $\mathcal{H}$ is a density operator with a Gaussian characteristic function, i.e.:
\begin{equation}
 \chi(\alpha)=\exp(\text{i}\boldsymbol{d}^T\boldsymbol{\alpha} - 
 \frac{1}{2}\boldsymbol{\alpha}^T\Sigma\boldsymbol{\alpha}),\qquad\boldsymbol{\alpha}=(\text{Re}\alpha_1,\hdots,\text{Re}\alpha_N,\text{Im}\alpha_1,\hdots,\text{Im}\alpha_N)^T\in \mathbb{R}^{2N}\,,
\end{equation}
with the displacement vector $\boldsymbol{d}\in\mathbb{R}^{2N}$ and the real, symmetric $2N\times 2N$ covariance matrix $\Sigma$. The displacement $\boldsymbol{d}$ of any Gaussian state can always be 
brought to zero by a local unitary transformation \cite{AdessoAndvancesPerspectives}. Therefore it does not contain any information about correlations between subsystems or unitarily invariant properties of the state, 
and will be set to zero in the following. A Gaussian state is then exclusively characterized by its covariance matrix $\Sigma$. The corresponding density operator of the state is given by \cite{QINFGaussStatesWang}:
\begin{equation}
 \rho_\Sigma=\frac{1}{\pi^N} \int\limits_{\mathbb{R}^{2N}}\exp(- \frac{1}{2}\boldsymbol{\alpha}^T\Sigma\boldsymbol{\alpha})  D_{-{\alpha}}\text{d}^{2N}\alpha\,.
 \label{eq:explGST}
\end{equation}
Any matrix $\Sigma$ is the covariance matrix of a Gaussian state if and only if \cite{SymplGeoGosson,GaussCVInfNapoli,AdessoAndvancesPerspectives,QINFGaussStatesWang}:\begin{equation}
\Sigma+\text{i}J\geq 0\,,
 \text{with the symplectic form } J=
\begin{pmatrix}
0 & \mathbb{1}\\
-\mathbb{1} & 0
\end{pmatrix}\,,
\end{equation}
where $\geq 0$ means positive semidefinite.
Gaussian unitary transformations $\mathcal{U}_G$ are unitary operators of the form:
\begin{equation}
 \mathcal{U}_G=\exp\big(-\text{i}\sum_{i,j=1}^N P_{ij}a_ia_j+P^*_{ij}a^\dagger_ia^\dagger_j+Q_{ij}a_i^\dagger a_j\big), \,\qquad P=P^T, Q=Q^\dagger\,.
\end{equation}
Together with global phase transformations the Gaussian unitary transformations form the class of unitary operators which transform Gaussian states $\rho_\Sigma$ into Gaussian states $\rho_{\tilde\Sigma}$ \cite{InvariantMeasuresPureGaussian}. 
The transformations $\mathcal{U}_G$ are a representation of the real symplectic group $\text{Sp}(2N,\mathbb{R})$, and lead to a symplectic transformation $S$ of the covariance 
matrix \cite{AdessoAndvancesPerspectives,GaussCVInfNapoli,InvariantMeasuresPureGaussian}:  
\begin{equation}
 \mathcal{U}_G\rho_\Sigma\mathcal{U}^\dagger_G=\rho_{S^T\Sigma S},\qquad\mbox{with}\; S^TJS=J\,.
\end{equation}
For any covariance matrix $\Sigma$ there exists a symplectic transformation $S$ such that $S^T\Sigma S$ is diagonal, with each diagonal entry appearing twice \cite{SymplGeoGosson}:
\begin{equation}
 S^T\Sigma S = D =
\begin{pmatrix}
\mathcal{N} & 0\\
0 & \mathcal{N}
\end{pmatrix},\qquad \mathcal{N}=\text{diag}(\nu_1,\hdots,\nu_N)\qquad \nu_i\geq 1\,.
\label{eq:D}
\end{equation}
The $N$ values $\nu_i$ are called symplectic eigenvalues of $\Sigma$. They characterize a Gaussian state up to unitary transformations, and are in this sense the equivalent to the eigenvalues of the 
density matrix in quantum systems with finite-dimensional Hilbert space. The density operator corresponding to the diagonal covariance matrix $D$ is a tensor product of thermal states of the harmonic 
oscillator \cite{AdessoAndvancesPerspectives}:
\begin{equation}
 \rho_D=\bigotimes\limits_{i=1}^N\frac{2}{\nu_i+1}\sum_{n=0}^\infty \left(\frac{\nu_i-1}{\nu_i+1}\right)^n\ket{n}_i\bra{n}_i\, ,
\end{equation}
where one may identify $\nu_i = \coth\left(\frac{\hbar\omega_i}{2{\rm k_B}T}\right)$.
This state is pure if and only if all $\nu_i$ are equal to one, that is if it is the vacuum state. Hence all pure Gaussian states (the set of squeezed states) 
are Gaussian unitary transformations of the vacuum state. 

\section{Hilbert-Schmidt measure for Gaussian states}\label{sec:HSmeasure}
\subsection{Hilbert-Schmidt volume element}\label{sec:HSvol}
In the literature a variety of different physically motivated measures on quantum state spaces is used. A thorough introduction to these measures for systems with a finite dimensional Hilbert space can be found in \cite{geometryQS}.
One important and widely used measure is the Hilbert-Schmidt measure. It is the measure induced by the Hilbert-Schmidt metric. The Hilbert-Schmidt distance of two states $\rho$ and $\rho'$ is defined as:
\begin{equation}
 d_{HS}(\rho,\rho')=\sqrt{\text{tr}\big((\rho-\rho')^2\big)}\,.
 \label{eq:HSdef}
\end{equation}
An important property of this metric is that it is invariant under unitary transformation of the density operators. For two Gaussian states $\rho_\Sigma$ and $\rho_{\Sigma'}$, the Hilbert-Schmidt distance can be calculated (see Appendix \ref{sec:APPHS}):
\begin{equation}
 d_{HS}(\rho_\Sigma,\rho_{\Sigma'})=\sqrt{\frac{1}{\sqrt{\det\Sigma}}+\frac{1}{\sqrt{\det\Sigma'}}-2\frac{1}{\sqrt{\det\frac{1}{2}(\Sigma+\Sigma')}}}\,.
 \label{eq:HSsigma}
\end{equation}
Since the metric is unitarily invariant, this expression should be symplectically invariant on the level of covariance matrices. This is indeed the case, because all symplectic matrices $S$ have a determinant of one \cite{SymplGeoGosson}:
\begin{equation}
 \det(S^T\Sigma S)=\det S^T\det\Sigma\det S=\det\Sigma\,.
\end{equation}
The infinitesimal distance element of the Hilbert-Schmidt metric takes the simple form:
\begin{equation}
\text{d}s^2_{HS}=\text{tr}\big((\text{d}\rho)^2\big)\,.
\end{equation}
If one restricts oneself to the manifold of Gaussian states, $\text{d}\rho$ becomes:
\begin{equation}
\begin{split}
\text{d}\rho_{\Sigma}&=\frac{1}{\pi^N} \int\limits_{\mathbb{R}^{2N}}\text{d}^{2N}\alpha D_{-\boldsymbol{\alpha}}\text{d}\big(\exp(- \frac{1}{2}\boldsymbol{\alpha}^T\Sigma\boldsymbol{\alpha})\big)\\
&=\frac{1}{\pi^N} \int\limits_{\mathbb{R}^{2N}}\text{d}^{2N}\alpha D_{-\boldsymbol{\alpha}}\exp(- \frac{1}{2}\boldsymbol{\alpha}^T\Sigma\boldsymbol{\alpha})\big(-\frac{1}{2}\boldsymbol{\alpha}^T\text{d}\Sigma\boldsymbol{\alpha}\big)\\
&=\frac{1}{\pi^N} \int\limits_{\mathbb{R}^{2N}}\text{d}^{2N}\alpha D_{-\boldsymbol{\alpha}}\exp(- \frac{1}{2}\boldsymbol{\alpha}^T\Sigma\boldsymbol{\alpha})\big(\exp(- \frac{1}{2}\boldsymbol{\alpha}^T\text{d}\Sigma\boldsymbol{\alpha})-1\big)=\rho_{\Sigma+\text{d}\Sigma}-\rho_{\Sigma}\,.
\end{split}
\end{equation}
The infinitesimal distance element for Gaussian states $\text{d}s_{HS}^2=d_{HS}(\rho_{\Sigma+\text{d}\Sigma},\rho_{\Sigma})^2$ is derived in Appendix \ref{sec:APPwegele1}:
\begin{equation}
 \text{d}s_{HS}^2=\frac{1}{16\sqrt{\det\Sigma}}\Big(\big(\text{tr}(\Sigma^{-1}\text{d} \Sigma)\big)^2+2\text{tr}\big((\Sigma^{-1}\text{d} \Sigma)^2\big)\Big)\,.
 \label{eq:wegeleallg}
\end{equation}
The volume element of the Hilbert-Schmidt measure can then be obtained by finding the explicit form of the metric tensor $g$:
\begin{equation}
 \text{d}s_{HS}^2=\sum_{\alpha,\beta}\text{d}\Sigma_{\alpha}g_{\alpha\beta}\text{d}\Sigma_{\beta} \qquad \text{d}V_{HS}=\sqrt{\det g}\prod_{\alpha}\text{d}\Sigma_{\alpha}\,,
 \label{eq:HSvolelebasic}
\end{equation}
where $\alpha$ and $\beta$ denote some indices labeling all matrix elements of $\text{d}\Sigma$.

\subsection{Distribution of symplectic eigenvalues}\label{sec:HSdist}
\begin{figure}[htp]
  \begin{center}
   \subfigure[$N=1$]{%
	\includegraphics[trim = 0px 0px 0px 0px,width=0.45\textwidth]{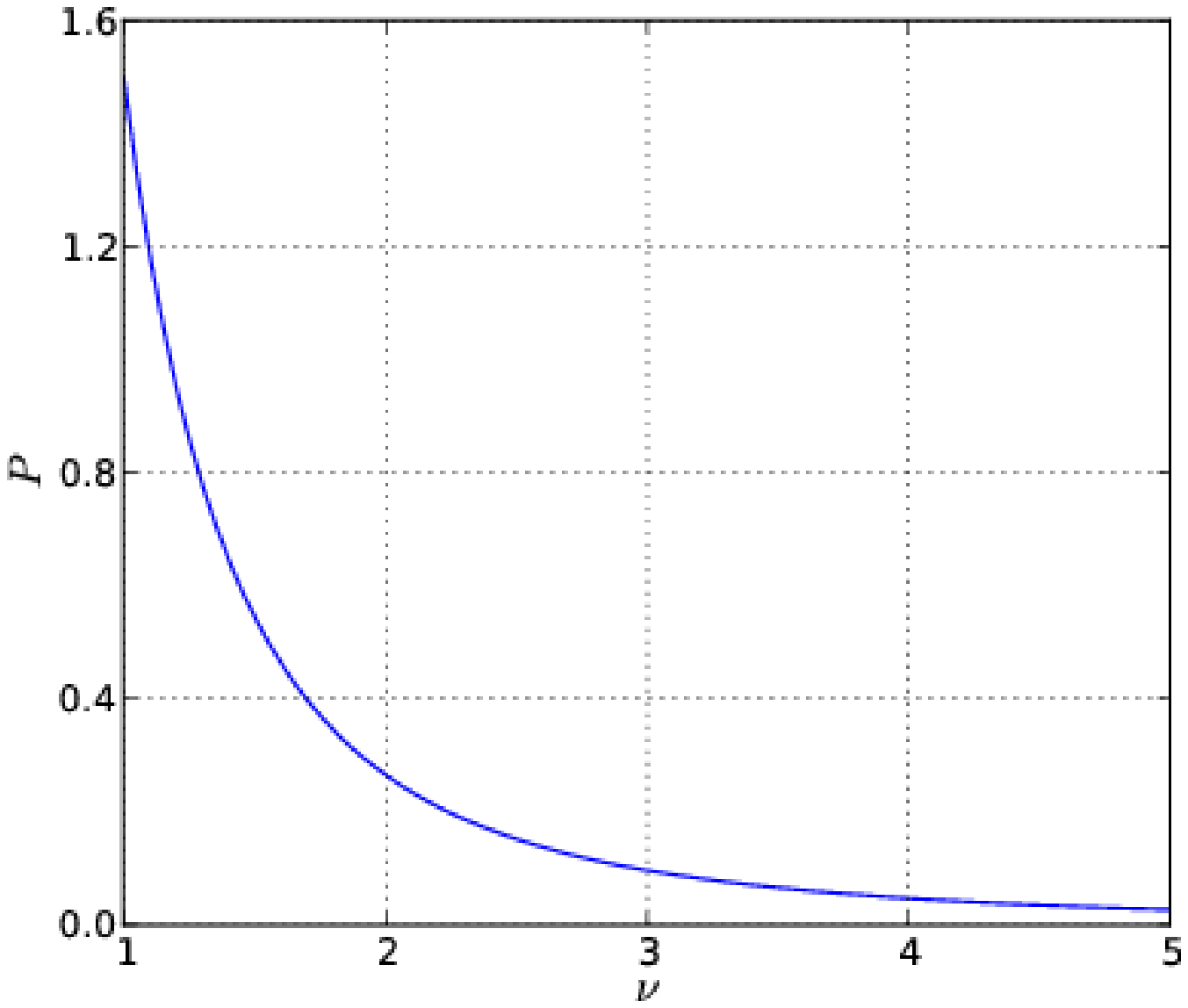}
   }%
 \hspace{0.02\textwidth}
   \subfigure[$N=2$]{%
\includegraphics[trim = 80px 50px 80px 50px,width=0.45\textwidth,clip=true]{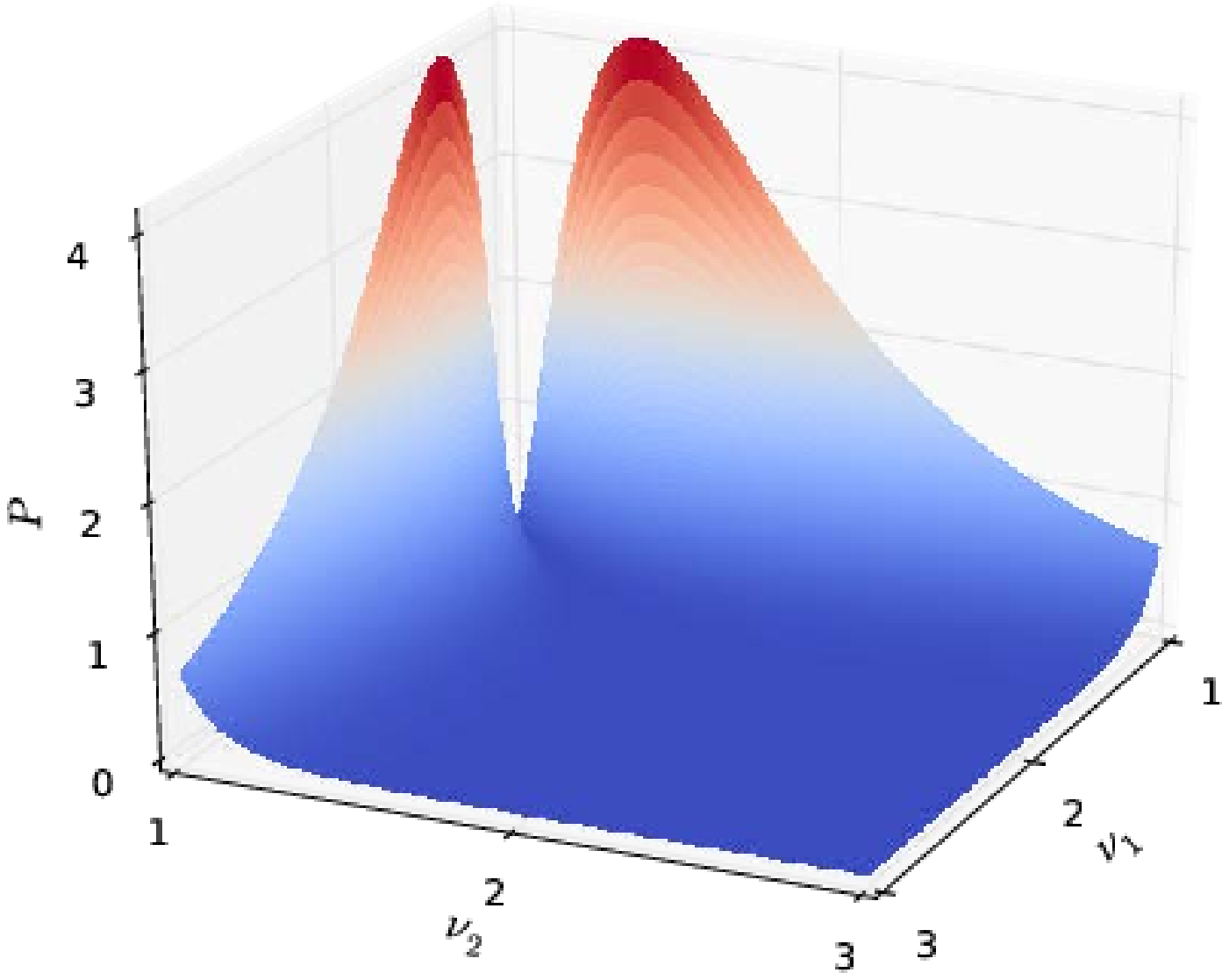}
   }
   \end{center}
 \caption{Normalized distribution of symplectic eigenvalues for Hilbert-Schmidt measure for one mode ($N=1$) and two mode ($N=2$) Gaussian states.}
      \label{fig:probdist}
 \end{figure}

For quantum states of systems with a finite-dimensional Hilbert space the distribution of the eigenvalues of density matrices for the Hilbert-Schmidt measure can be derived \cite{HallRandQC}. 
In the same way it is possible to derive the distribution of symplectic eigenvalues $\{\nu_i|i=1,\hdots,N\}$ of Gaussian states for this measure.\\
Any covariance matrix $\Sigma$ of a Gaussian state can be written as a symplectic transformation of a diagonal matrix $D$ (see equation (\ref{eq:D})). One may write:
\begin{equation}
 \Sigma+\text{d} \Sigma = S^TDS+\text{d} \Sigma=S^T(D+\delta D)S\,.
 \label{eq:dsigma}
\end{equation}
The expression $D+\delta D$ can be decomposed as an infinitesimal shift in the symplectic eigenvalues followed by an infinitesimal symplectic transformation:
\begin{equation}
 D+\delta D=(\mathbb{1}+\text{d} X^T)(D+\text{d}D)(\mathbb{1}+\text{d} X)=D+\text{d}D+\text{d} X^T D+D\text{d} X\,.
\end{equation}
$X$ is the generator of a symplectic transformation and is therefore a Hamiltonian matrix $(JX)^T=JX$ \cite{LIE}. 
Inserting (\ref{eq:dsigma}) in (\ref{eq:wegeleallg}) gives (see Appendix \ref{sec:APPwegele2}):
\begin{equation}
\begin{split}
 \text{d}s^2_{HS}=&\frac{1}{16\sqrt{\det D}}\Big(\big(\text{tr}(D^{-1}\text{d}D)\big)^2+2\text{tr}\big((D^{-1}\text{d} D)^2\big)\\
 &+4\text{tr}\big((\text{d}X)^2\big)+4\text{tr}(\text{d}XD^{-1}\text{d}X^TD)\Big)\,.
 \end{split}
 \label{eq:wegelediag}
\end{equation}
Since no terms of the form $\text{d}D\text{d}X$ appear, the metric tensor is of block form $g=g^{D}\oplus g^{S}$, with $g^D$ corresponding to the symplectic eigenvalues and $g^S$ to the symplectic transformation.\\
A general symplectic transformation $S\in\text{Sp}(2N,\mathbb{R})$ is of dimension $N(2N+1)$. Together with the $N$ symplectic eigenvalues, the expression $S^TDS=\Sigma$ is $N$-fold overdetermined, since the dimension of $\Sigma$ is just $N(2N+1)$. Thus the symplectic transformation $S$ which transforms $D$ to $\Sigma$ is not unique. The symplectic transformation:
\begin{equation}
 \tilde S = QS, \qquad Q=
 \begin{pmatrix}
 Q_1&Q_2\\
 Q_3&Q_4
 \end{pmatrix}, \qquad Q^TQ=\mathbb{1}, \qquad Q_i \text{ diagonal}
 \label{eq:eqivS}
\end{equation}
also transforms $D$ to $\Sigma=S^TDS$, because $D$ and $Q$ commute:
\begin{equation}
\begin{split}
 \tilde S^T D\tilde S=S^TQ^TDQS&=S^T 
 Q^TQ DS
 =S^TDS\,.
 \end{split}
\end{equation}
The transformations $S$ and $\tilde S$ are in this sense equivalent. The matrices $Q$ form a group that is the direct sum of $N$ rotation groups $\text{SO}(2)$, and are therefore of dimension $N$. 
Hence they are the only freedom in the choice of $S$. In order to derive the Hilbert-Schmidt volume element explicitly, a $2N^2$-dimensional parameterization of infinitesimal symplectic transformations has to be found, 
in which this freedom is not included. \\Any Hamiltonian matrix $X$ can be written as \cite{LIE}:
\begin{equation}
X=
 \begin{pmatrix}
A & B\\
C & -A^T
\end{pmatrix}
\qquad B=B^T \qquad C=C^T\,.
\end{equation}
And the generator of the matrices $Q$ is:
\begin{equation}
 R=-R^T=
 \begin{pmatrix}
  0&G\\
  -G&0
 \end{pmatrix}
 \qquad G \text{ diagonal}.
\end{equation}
For a given infinitesimal transformation $\tilde S = \mathbb{1}+\text{d}\tilde X$, consider the equivalent transformation:
\begin{equation}
 \mathbb{1}+\text{d}X=(\mathbb{1}+\text{d} R)\tilde S=\mathbb{1}+\text{d}R+\text{d}\tilde X=\mathbb{1}+
 \begin{pmatrix}
0&\text{d}G\\
-\text{d}G&0
\end{pmatrix}+
\begin{pmatrix}
\text{d}\tilde A&\text{d}\tilde B\\
\text{d}\tilde C&-\text{d}\tilde A^T
\end{pmatrix}\,.
\end{equation}
Choosing $\text{d}G=\text{diag}(\text{d}c_{11},\hdots,\text{d}c_{NN})$ and assigning $\text{d}b_{ii}=\text{d}(\tilde b_{ii}+\delta_{ij}c_{ii})$ gives a suitable $2N^2$-dimensional parameterization of $\text{d}X$:
\begin{equation}
 \text{d}X=
 \begin{pmatrix}
  \text{d}A&\text{d}B\\
  \text{d}C&-\text{d}A^T
 \end{pmatrix}
 \qquad \text{d}B=\text{d}B^T
 \qquad \text{d}C=\text{d}C^T
 \qquad (\text{d}C)_{ii}=0\,;\, i=1,\hdots,N\,.
 \label{eq:sparam}
\end{equation}
Thus the volume element $\text{d}V_{HS}$ becomes:
\begin{equation}
 \text{d}V_{HS}=\sqrt{\det g} \Big(\prod\limits_{i=1}^{N}\text{d}\nu_i\Big)\Big(\prod\limits_{l,m=1}^N\text{d}a_{lm}\Big)\Big(\prod\limits_{r\geq s=1}^N\text{d}b_{rs}\Big)\Big(\prod_{q>p=1}^N\text{d}c_{qp}\Big)\,.
 \label{eq:volelehs}
\end{equation}
The metric tensor $g$ is derived in Appendix \ref{sec:APPexplTens}. The result is:
\begin{equation}
 \sqrt{\det g}=\frac{\sqrt{N+1}}{4^{N^2}}\Big(\frac{1}{\prod_{k=1}^N\nu_k}\Big)^{N(N+\frac{5}{2})-1}
 \prod\limits_{l>m=1}^N(\nu_l^2-\nu_m^2)^2\,.
 \label{eq:explmeasure}
\end{equation}
The integral over the symplectic Group is divergent, since the symplectic group is not a compact group. Hence the Hilbert-Schmidt volume of the space of mixed Gaussian states is infinite. 
However, the integral over the symplectic group only contributes as a trivial factor, since it is independent of the integrand $\sqrt{\det g}$. Integrating over the symplectic eigenvalues gives a finite value for all $N$, so that the distribution of symplectic eigenvalues $P_N(\nu_1,\hdots,\nu_N)\propto\sqrt{\det g}$ is normalizable. For the simplest cases $N=1$ and $N=2$ one obtains:
\begin{equation}
 P_1(\nu)=\frac{3}{2}\nu^{-\frac{5}{2}} \qquad P_2(\nu_1,\nu_2)=\frac{525}{8}\frac{(\nu_1^2-\nu_2^2)^2}{\nu_1^8\,\nu_2^8}
 \qquad\,\,\,\, \nu_i\in[1,\infty)\,.
\end{equation}
These distributions are illustrated in figure \ref{fig:probdist}. In general, large values of symplectic eigenvalues occur with a small probability (Hilbert-Schmidt measure) - pure states are more likely. This is nicely in
line with the corresponding findings for qubit-systems (see \cite{geometryQS}). Even the node structures (e.g. zero along $\nu_1=\nu_2$ for $N=2$), known from qubits, are reflected. 
For $N=1$ the maximum of the distribution is located at the pure states $\nu=1$. This changes for $N\geq 2$, because the distribution is zero when two symplectic eigenvalues are equal. 
For the case $N=2$ the maximum is at  $\nu_{1(2)}=1,\nu_{2(1)}=\sqrt{2}$.

\subsection{Distribution of purity and entropy}\label{sec:HSprop}
\begin{figure}[htp]
  \begin{center}
   \subfigure[Mean purity]{%
	\includegraphics[trim = 0px 0px 0px 0px,width=0.45\textwidth]{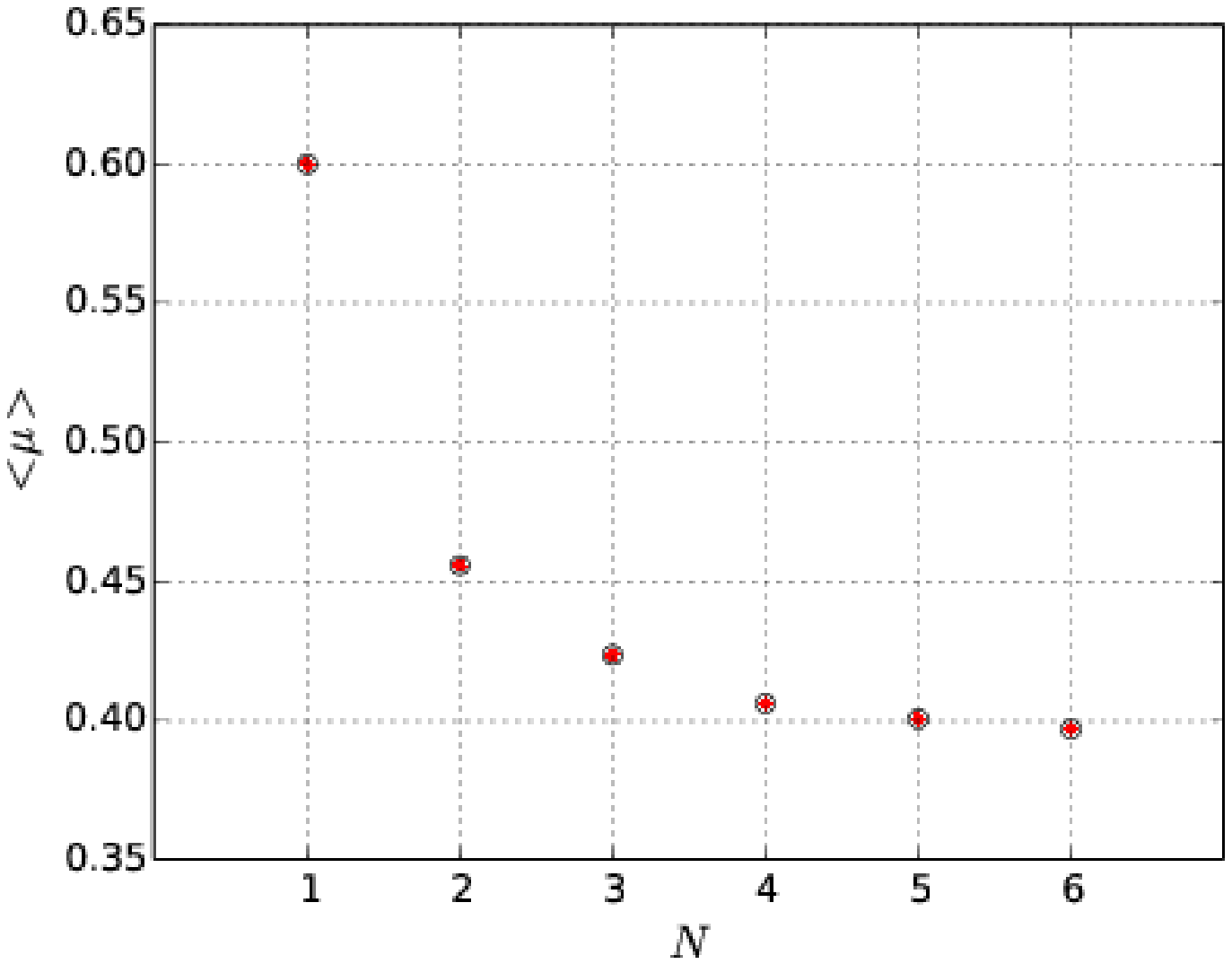}
   }%
 \hspace{0.02\textwidth}
   \subfigure[Mean von Neumann entropy]{%
\includegraphics[trim = 0px 0px 0px 0px,width=0.45\textwidth,clip=true]{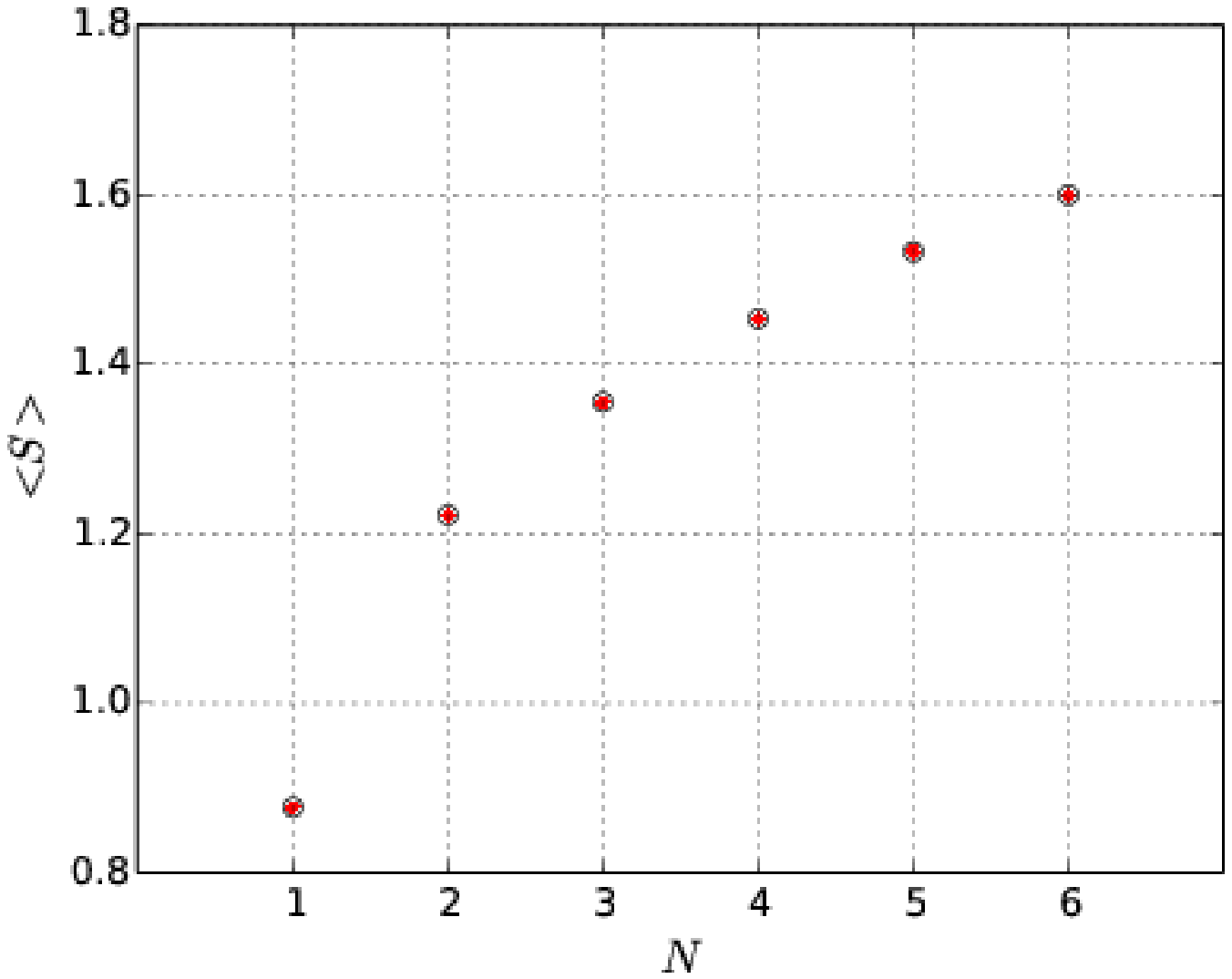}
   }
   \end{center}
 \caption{Average purity (a) and average von-Neumann entropy (b) of a Gaussian state (Hilbert-Schmidt ensemble) as a function of the number $N$ of modes.}
      \label{fig:means}
 \end{figure}

\begin{figure}[htp]
\centering
\begin{subfigure}
\centering
\includegraphics[trim = 0px 0px 0px 0px,width=0.45\textwidth]{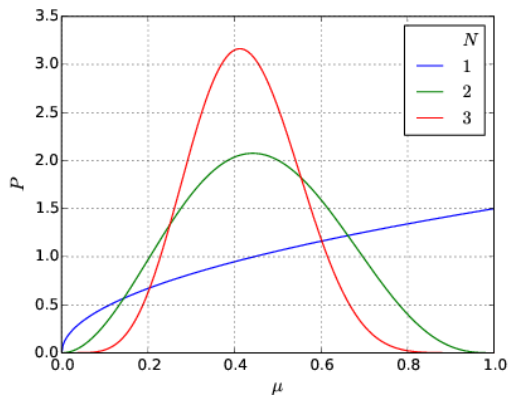}
\end{subfigure}
\caption{Distribution of the purity of Gaussian states in the Hilbert-Schmidt ensemble for mode numbers $N=1,2,3$.}%
\label{fig:mudist}%
\end{figure}

With the help of the distribution (\ref{eq:volelehs}) it is possible to calculate expectation values of all unitarily invariant properties of $N$-mode Gaussian states for the Hilbert-Schmidt measure, 
as all these properties can be expressed solely by the symplectic eigenvalues. From the derivation of the Hilbert-Schmidt distance of two Gaussian states (Appendix \ref{sec:APPHS}) one immediately sees that 
the purity $\mu=$tr$(\rho^2)$ of a Gaussian state is:
\begin{equation}
\mu(\rho_\Sigma)=\text{tr}\big((\rho_\Sigma)^2\big)=\frac{1}{\sqrt{\det\Sigma}}=\frac{1}{\prod_{i=1}^N\nu_i}\,.
\end{equation}
A second important quantity, the von Neumann entropy $S$ of a Gaussian state, is given by \cite{entropyGauss}:
\begin{equation}
S(\rho_\Sigma)=-\text{tr}(\rho_\Sigma\ln \rho_\Sigma)=\sum_{i=1}^N\left\{ \frac{\nu_i+1}{2}\ln\Big(\frac{\nu_i+1}{2}\Big)-\frac{\nu_i-1}{2}\ln\Big(\frac{\nu_i-1}{2}\Big)\right\}\,.
\end{equation}
For low mode numbers $N$ the mean values of these quantities can easily be calculated numerically with high precision using the derived volume element. The results are shown in figure \ref{fig:means}. In general, Gaussian states in the Hilbert-Schmidt ensemble become more mixed for high mode numbers, since the purity decreases and the von Neumann entropy increases as $N$ increases. This is also the case for many-qubit systems, where the mean purity has been computed analytically in Hilbert-Schmidt measure as a function of the Hilbert-space dimension. The results can be found in \cite{PurityQubit}(figure 2(a)), and can be compared to figure \ref{fig:means}. \\
 It is also possible to obtain the entire distribution of the purity of Gaussian states in Hilbert-Schmidt measure analytically (see Appendix \ref{sec:APPdist}). 
 For the more complicated von Neumann entropy such an expression cannot be given analytically. For $N=1$ the purity is simply distributed like $\sqrt{\mu}$. 
 This distribution and the ones for $N=2$ and $N=3$ modes are shown in figure \ref{fig:mudist}. The variance of the distributions is decreasing as $N$ increases. 
 It is important to note that as a consequence of the infinite Hilbert-space dimension the purity can be zero, however the Hilbert-Schmidt ensemble does not contain infinitely mixed states, so that the distribution of the purity is zero for $\mu=0$. 

\section{Bures measure for single mode Gaussian states}\label{sec:BURES}
The Bures distance of two quantum states can be defined as a function of the fidelity $F$:
\begin{equation}
d_B(\rho,\rho')=\sqrt{2\big(1-\sqrt{F(\rho,\rho')})\big)}\qquad F(\rho,\rho')=\Big(\text{tr}\big(\sqrt{\sqrt{\rho}\rho'\sqrt{\rho}}\big)\Big)^2\,.
\end{equation}
An explicit formula for the fidelity of two $N$-mode Gaussian states has been derived in \cite{FidelityMMGauss}. For $N=1$ this expression simplifies to:
\begin{equation}
F(\rho_\Sigma,\rho_{\Sigma'})=\frac{2}{\sqrt{\det(\Sigma+\Sigma')+\mathcal{P}}-\sqrt{\mathcal{P}}}\qquad\mathcal{P}=(\det\Sigma-1)(\det\Sigma'-1)\,.
\end{equation}
Following the same steps as for the Hilbert-Schmidt measure, one can calculate the infinitesimal volume element of the Bures measure in the $N=1$ case. For $N>1$ this is not possible offhand, due to the complicated structure of the general formula for the fidelity. \\
Using the notation of the last section, the result is:
\begin{equation}
\begin{split}
\text{d}V_B
=\frac{1}{4}\frac{\nu^2}{\nu^2+1}\frac{1}{\sqrt{\nu^2-1}}\text{d}\nu\,\text{d}a\,\text{d}b
\end{split}\,.
\label{eq:BURESvele}
\end{equation}
In Bures measure the distribution of symplectic eigenvalues is not normalizable, as it only decreases with $1/\nu$ for large $\nu$. Thus the Bures ensemble is much more concentrated on mixed states than the Hilbert-Schmidt ensemble. One can define the mean value of a quantity $f(\nu)$ with respect to the not normalizable Bures distribution in the following natural way:
\begin{equation}
\langle f \rangle_B\,=\lim\limits_{\nu_m\rightarrow\infty}\frac{\int_{1}^{\nu_m}f(\nu)\tilde P_B(\nu)\text{d}\nu}{\int_{1}^{\nu_m}\tilde P_B(\nu)\text{d}\nu}\qquad \tilde P_B(\nu)=\frac{\nu^2}{\nu^2+1}\frac{1}{\sqrt{\nu^2-1}}\,.
\end{equation}
It is easy to see that with this definition the mean purity of one-mode Gaussian states in the Bures 
ensemble $\braket{\mu}_B$ is zero, and the mean von Neumann entropy $\braket{S}_B$ infinite. Thus when no further restriction on the states is made (e.g. fixed energy), the Bures ensemble of Gaussian states contains almost only infinitely mixed states. It is not clear whether this still holds for $N>1$. However, a consequence might be that the probability to find an entangled Gaussian state in the Bures distribution is zero, as numerical calculations suggest for $N=2$ \cite{SlaterSep}.

\section{Summary and Outlook}\label{sec:SUM}
In contrast to systems with finite dimensional Hilbert space, much less is known about the geometry of quantum state spaces in continuous 
variable systems. Gaussian states allow for an easy access to this topic because of their simple structure. In this work the 
Hilbert-Schmidt measure has been studied on the domain of mixed Gaussian states. It turns out that in spite of the fact that the manifold of 
Gaussian states is unbounded, Gaussian states distributed uniformly according to the Hilbert-Schmidt measure (\ref{eq:volelehs}) have 
a finite mean von Neumann entropy and a nonzero mean purity, which is reflected in the normalizability of the distribution of symplectic 
eigenvalues. This is not obvious, as it turns out that for an ensemble of one mode Gaussian states for the Bures measure (\ref{eq:BURESvele}) 
this distribution cannot be normalized, and the ensemble consists of almost only infinitely mixed states. It is not clear whether this is 
also true for the Bures ensemble of multi mode Gaussian states.\\
With the distribution of symplectic eigenvalues at hand, similar to the distribution of eigenvalues in qubit systems, all unitarily 
invariant properties of the ensemble can be calculated. As examples,
we study the distribution of purity and the von-Neumann entropy of Gaussian quantum states. Entanglement does not belong to these unitarily 
invariant properties. However, the Hilbert-Schmidt volume element (\ref{eq:HSvolelebasic}) can be used straightforwardly for the study 
of entanglement in multi mode Gaussian quantum states. For any compact subset of two mode Gaussian states (for example the setcounterof two-mode
states for fixed energy expectation value), the probability for a separable state can be calculated numerically for the Hilbert-Schmidt measure 
using the analytical form of the volume element and the Simon-Peres–Horodecki separability criterion \cite{PPTcrit}. In addition, it would 
also be possible to consider conditioned bipartite Gaussian states and study "conditioned volumes". In this area some new results for the 
two qubit system have been obtained recently \cite{Simon}, and it would be interesting to see whether some of those results still hold for 
Gaussian states.

\section*{Acknowledgments}

It is a pleasure to thank Simon Milz and Karol Życzkowski for discussions and advice. We are grateful to Simon 
for a careful reading of the manuscript, too.


\begin{appendix}
\section{Hilbert-Schmidt metric for Gaussian states}\label{sec:APPHS}
The expression (\ref{eq:HSdef}) is to be calculated for Gaussian states.
\begin{equation}
 \begin{split}
  &\text{tr}\big((\rho_\Sigma-\rho_{\Sigma'})^2\big) = \text{tr}\big((\rho_\Sigma)^2\big)+\text{tr}\big((\rho_{\Sigma'})^2\big)-2\text{tr}\big(\rho_\Sigma\rho_ {\Sigma'}\big)
 \nonumber
 \end{split}
\end{equation}
The characteristic function of the product of two Gaussian states has been calculated in \citep{FidelityMMGauss}. The trace is given by evaluating the characteristic function at zero. One obtains:
\begin{equation}
\text{tr}\big(\rho_\Sigma\rho_ {\Sigma'}\big)=\frac{1}{\sqrt{\det\frac{1}{2}(\Sigma+\Sigma')}}\,,
\nonumber
\end{equation}
from which equation (\ref{eq:HSsigma}) follows immediately.

\section{Distance element for the Hilbert-Schmidt metric}\label{sec:APPwegele1}
The infinitesimal distance element for Gaussian states is:
\begin{equation}
 \begin{split}
 \text{d}s_{HS}^2&=d_{HS}(\rho_{\Sigma+\text{d}\Sigma},\rho_{\Sigma})^2=\big(\det\Sigma\big)^{-\frac{1}{2}}+\big(\det(\Sigma+\text{d} \Sigma)\big)^{-\frac{1}{2}}-2\big(\det(\Sigma+\frac{\text{d} \Sigma}{2})\big)^{-\frac{1}{2}}\\
 &=\big(\det\Sigma\big)^{-\frac{1}{2}}\Big(1+\big(\det(\mathbb{1}+\Sigma^{-1}\text{d} \Sigma)\big)^{-\frac{1}{2}}-2\big(\det(\mathbb{1}+\Sigma^{-1}\frac{\text{d} \Sigma}{2})\big)^{-\frac{1}{2}}\Big)\,.
 \end{split}
 \nonumber
\end{equation}
With use of the relation $\det (M) = \exp\big(\text{tr}(\ln (M))\big)$ \cite{MatrixAna} and the series representation of the matrix 
logarithm. Up to second order in $\text{d} \Sigma$ one obtains:
\begin{equation}
\begin{split}
 \big(\det(\mathbb{1}+\Sigma^{-1}\text{d} \Sigma)\big)^{-\frac{1}{2}}&=\exp\big(-\frac{1}{2}\text{tr}(\ln(\mathbb{1}+\Sigma^{-1}\text{d} \Sigma))\big)
 \\&=1-\frac{1}{2}\text{tr}(\Sigma^{-1}\text{d} \Sigma)+\frac{1}{4}\text{tr}\big((\Sigma^{-1}\text{d} \Sigma)^2\big)+\frac{1}{8}\big(\text{tr}(\Sigma^{-1}\text{d} \Sigma)\big)^2\,.
 \end{split}
 \nonumber
\end{equation}
The result (\ref{eq:wegeleallg}) follows from the same calculation for the second determinant.

\section{Distance element for the symplectic eigenvalue decomposition}\label{sec:APPwegele2}
Using (\ref{eq:dsigma}) one obtains for the two terms in (\ref{eq:wegeleallg}):
\begin{equation}
 \begin{split}
  &\text{tr}(\Sigma^{-1}\text{d} \Sigma)=\,\text{tr}(D^{-1}\text{d} D)+2\text{tr}(\text{d}X),\\
  &\text{tr}\big((\Sigma^{-1}\text{d} \Sigma)^2\big)=\,\text{tr}\big((D^{-1}\text{d} D)^2\big)+2\text{tr}\big((\text{d}X)^2\big)
  +2\text{tr}(D^{-1}\text{d}X^T S^TD\text{d}X)\\
  &\qquad\qquad\qquad\quad+4\text{tr}(\text{d} D\text{d} X D^{-1})\,.
 \end{split}
\nonumber
\end{equation}
From $J^2=-\mathbb{1}$ and $(J\text{d}X)^T=J\text{d}X$ we find:
\begin{equation}
 \begin{split}
  \text{tr}(\text{d} X)&=-\text{tr}(J^2\text{d} X)=-\text{tr}(J(J\text{d} X)^T)=-\text{tr}(J\text{d} X^TJ^T)=-\text{tr}(\text{d} X)=0\,.
 \end{split}
 \nonumber
\end{equation}
In addition $[J,D]=[J,D^{-1}]=[J,\text{d} D]=[D,\text{d} D]=0$ and $D^T=D$ applies, and therefore:
\begin{equation}
\begin{split}
 \text{tr}(\text{d} D\text{d} X D^{-1})&=-\text{tr}(\text{d} DJ^2\text{d} X D^{-1})
 =-\text{tr}(\text{d} D J(J\text{d} X)^T D^{-1})\\
 &=-\text{tr}(\text{d} DJ\text{d} X^T J^TD^{-1})=\text{tr}(\text{d} DJ^2\text{d} X D^{-1})=
 -\text{tr}(\text{d} D\text{d} X D^{-1})=0\,.
 \end{split}
 \nonumber
\end{equation}
Thus all terms proportional to $\text{d} D\text{d} X$ vanish. Overall one obtains the result (\ref{eq:wegelediag}).

\section{Explicit form of the metric tensor} \label{sec:APPexplTens}
The metric tensor $g^D$ arises from the terms in (\ref{eq:wegelediag}) containing $\text{d}D$. Explicitly, they are given by:
\begin{equation}
 \begin{split}
  &\big(\text{tr}(D^{-1}\text{d}D)\big)^2=4\sum_{i,j=1}^N\frac{\text{d}\nu_i\text{d}\nu_j}{\nu_i\nu_j}\qquad 2\text{tr}\big((D^{-1}\text{d}D)^2\big)=4\sum_{i=1}^N\frac{\text{d}\nu_i^2}{\nu_i^2}\,.
 \end{split}
 \nonumber
\end{equation}
Thus the explicit form of $g^D$ reads:
\begin{equation}
 g^D_{ij}=\frac{1}{4\sqrt{\det D}}\frac{(1+\delta_{ij})}{\nu_i\nu_j}=\frac{1}{4\prod_{k=1}^N\nu_k}\frac{(1+\delta_{ij})}{\nu_i\nu_j}\,.
 \nonumber
\end{equation}
Using the parameterization (\ref{eq:sparam}), the terms containing $\text{d}X$ give:
\begin{equation}
\begin{split}
 &\text{tr}\big((\text{d}X)^2\big)=2\text{tr}\big((\text{d}A)^2\big)+2\text{tr}(\text{d}B\text{d}C)=2\sum_{i,j=1}^N\text{d}a_{ij}\text{d}a_{ji}+2\sum_{i>j=1}^N(\text{d}b_{ij}\text{d}c_{ij}+\text{d}c_{ij}\text{d}b_{ij})
\end{split}
\nonumber
 \end{equation}
 \begin{equation}
\begin{split}
 &\text{tr}(\text{d}XD^{-1}\text{d}X^TD)\\
 &= \text{tr}(\text{d}A\mathcal{N}^{-1}\text{d}A^T\mathcal{N})+\text{tr}(\text{d}A^T\mathcal{N}^{-1}\text{d}A\mathcal{N})+\text{tr}(\text{d}B\mathcal{N}^{-1}\text{d}B\mathcal{N})+\text{tr}(\text{d}C\mathcal{N}^{-1}\text{d}C\mathcal{N})\\
 &=\sum_{i,j=1}^N\text{d}a_{ij}^2\Big(\frac{\nu_i}{\nu_j}+\frac{\nu_j}{\nu_i}\Big)+\sum_{i>j=1}^N(\text{d}b_{ij}^2+\text{d}c_{ij}^2)\Big(\frac{\nu_i}{\nu_j}+\frac{\nu_j}{\nu_i}\Big)+\sum_{i=1}^N\text{d}b_{ii}^2\,.
\end{split}
\nonumber
 \end{equation}
Overall, the distance element can be written as:
\begin{equation}
\begin{split}
(\text{d}s_{HS})^2=&
\begin{pmatrix}
\text{d}\nu_1\\
\vdots\\
\text{d}\nu_N
\end{pmatrix}^T
g^D
\begin{pmatrix}
\text{d}\nu_1\\
\vdots\\
\text{d}\nu_N
\end{pmatrix}
+ \frac{1}{4\prod_{k=1}^N\nu_k}\Big(\begin{pmatrix}
\text{d}a_{11}\\
\vdots\\
\text{d}a_{NN}
\end{pmatrix}^T
4\,\mathbb{1}
\begin{pmatrix}
\text{d}a_{11}\\
\vdots\\
\text{d}a_{NN}
\end{pmatrix}\\&
+\sum_{i>j=1}^N
\begin{pmatrix}
  \text{d}a_{ij}\\
  \text{d}a_{ji}
 \end{pmatrix}^T
 \begin{pmatrix}
  \frac{\nu_i}{\nu_j}+\frac{\nu_j}{\nu_i}&2\\
  2&\frac{\nu_i}{\nu_j}+\frac{\nu_j}{\nu_i}
 \end{pmatrix}
\begin{pmatrix}
  \text{d}a_{ij}\\
  \text{d}a_{ji}
 \end{pmatrix}\\&
+\begin{pmatrix}
\text{d}b_{11}\\
\vdots\\
\text{d}b_{NN}
\end{pmatrix}^T
\mathbb{1}
\begin{pmatrix}
\text{d}b_{11}\\
\vdots\\
\text{d}b_{NN}
\end{pmatrix}
+\sum_{i>j=1}^N
\begin{pmatrix}
  \text{d}b_{ij}\\
  \text{d}c_{ij}
 \end{pmatrix}^T
 \begin{pmatrix}
  \frac{\nu_i}{\nu_j}+\frac{\nu_j}{\nu_i}&2\\
  2&\frac{\nu_i}{\nu_j}+\frac{\nu_j}{\nu_i}
 \end{pmatrix}
\begin{pmatrix}
  \text{d}b_{ij}\\
  \text{d}c_{ij}
 \end{pmatrix}
\Big)\,.
\end{split}
\nonumber
\end{equation}
One can read the explicit form of the metric tensor. The measure $\sqrt{\det g}$ turns out to be:
\begin{equation}
\begin{split}
  \sqrt{\det g}&=\frac{1}{4^{N^2}}\Big(\frac{1}{\prod_{k=1}^N\nu_k}\Big)^{N(N+\frac{1}{2})}\sqrt{\det \big(\frac{1+\delta_{ij}}{\nu_i\nu_j}\big)}
 \prod\limits_{l>m=1}^N\Big(\big(\frac{\nu_i}{\nu_j}+\frac{\nu_j}{\nu_i}\big)^2-4\Big)\\
 &=\frac{1}{4^{N^2}}\Big(\frac{1}{\prod_{k=1}^N\nu_k}\Big)^{N(N+\frac{1}{2})}\sqrt{\frac{1}{\prod_{k=1}^N\nu_k^2}\det \big(1+\delta_{ij}\big)} \Big(\frac{1}{\prod_{k=1}^N\nu_k}\Big)^{2(N-1)}\prod\limits_{l>m=1}^N(\nu_l^2-\nu_m^2)^2\\
 &=\frac{1}{4^{N^2}}\Big(\frac{1}{\prod_{k=1}^N\nu_k}\Big)^{N(N+\frac{5}{2})-1}\sqrt{N+1}
 \prod\limits_{l>m=1}^N(\nu_l^2-\nu_m^2)^2\,.
 \label{eq:explmeasure2}
\end{split}
\nonumber
\end{equation}
\section{Distribution of purity}\label{sec:APPdist}
The distribution of the purity for mode number $N$ is given by:
\begin{equation}
P(\mu)=\int\limits_1^\infty \text{d}\nu_1...\int\limits_1^\infty \text{d}\nu_N P_N(\nu_1,...,\nu_N)\delta(\mu-\prod_{i=1}^N\frac{1}{\nu_i})\,.\nonumber
\end{equation}
Evaluating the $\nu_1$ integral with the delta function gives:
\begin{equation}
P(\mu)=\int\limits_1^{\mu^{-1}} \text{d}\nu_2\int\limits_1^{(\mu\nu_2)^{-1}} \text{d}\nu_3...\int\limits_1^{(\mu\nu_2...\nu_{N-1})^{-1}} \text{d}\nu_N P_N(\frac{1}{\mu\nu_2...\nu_N},\nu_2,...,\nu_N)\frac{1}{\mu^2\nu_2...\nu_N}\,.
\nonumber
\end{equation}
For any given $N$ the integrals can easily be computed analytically.

\end{appendix}


\end{document}